\documentclass[%
 reprint,
 amsmath,amssymb,
 superscriptaddress,
 aps
]{revtex4-2}

\usepackage{graphicx}
\usepackage{dcolumn}
\usepackage{bm}
\usepackage{here}
\usepackage{enumerate}
\usepackage{mathtools}
\usepackage{physics}

\makeatletter
\let\MYcaption\@makecaption
\makeatother
\usepackage{subcaption}
\makeatletter
\let\@makecaption\MYcaption
\makeatother

\usepackage{hyperref}
\begin{document}

\title{Comparative study of 1D and 2D convolutional neural network models with attribution analysis for gravitational wave detection from compact binary coalescences}

\author{Seiya Sasaoka}
\affiliation{Department of Physics, Tokyo Institute of Technology, 2-12-1 Ookayama, Meguro-ku, Tokyo 152-8551, Japan}%

\author{Naoki Koyama}
\affiliation{Graduate School of Science and Technology, Niigata University, 8050 Ikarashi-2-no-cho, Nishi-ku, Niigata City, Niigata 950-2181, Japan}%

\author{Diego Dominguez}
\affiliation{Department of Physics, Tokyo Institute of Technology, 2-12-1 Ookayama, Meguro-ku, Tokyo 152-8551, Japan}%

\author{Yusuke Sakai}
\affiliation{Department of Design and Data Science and Research Center for Space Science, Advanced Research Laboratories, \\ Tokyo City University, 3-3-1 Ushikubo-Nishi, Tsuzuki-ku, Yokohama, Kanagawa 224-8551, Japan}%

\author{Kentaro Somiya}
\affiliation{Department of Physics, Tokyo Institute of Technology, 2-12-1 Ookayama, Meguro-ku, Tokyo 152-8551, Japan}%

\author{Yuto Omae}
\affiliation{Artificial Intelligence Research Center, College of Industrial Technology, Nihon University, 1-2-1 Izumi-cho, Narashino, Chiba 275-8575, Japan}%

\author{Hirotaka Takahashi}
\affiliation{Department of Design and Data Science and Research Center for Space Science, Advanced Research Laboratories, \\ Tokyo City University, 3-3-1 Ushikubo-Nishi, Tsuzuki-ku, Yokohama, Kanagawa 224-8551, Japan}%
\affiliation{Institute for Cosmic Ray Research (ICRR), The University of Tokyo, 5-1-5 Kashiwa-no-Ha, Kashiwa City, Chiba 277-8582, Japan}
\affiliation{Earthquake Research Institute, The University of Tokyo, 1-1-1 Yayoi, Bunkyo-ku, Tokyo 113-0032, Japan}

\date{\today}

\begin{abstract}
Recent advancements in gravitational wave astronomy have seen the application of convolutional neural networks (CNNs) in signal detection from compact binary coalescences. This study presents a comparative analysis of two CNN architectures: one-dimensional (1D) and two-dimensional (2D) along with an ensemble model combining both. We trained these models to detect gravitational wave signals from binary black hole (BBH) mergers, neutron star-black hole (NSBH) mergers, and binary neutron star (BNS) mergers within real detector noise. Our investigation entailed a comprehensive evaluation of the detection performance of each model type across different signal classes. To understand the models' decision-making processes, we employed feature map visualization and attribution analysis. The findings revealed that while the 1D model showed superior performance in detecting BBH signals, the 2D model excelled in identifying NSBH and BNS signals. Notably, the ensemble model outperformed both individual models across all signal types, demonstrating enhanced detection capabilities. Additionally, input feature visualization indicated distinct areas of focus in the data for the 1D and 2D models, emphasizing the effectiveness of their combination.
\end{abstract}

\maketitle

\section{Introduction\label{sec:intro}}
The era of gravitational wave (GW) astronomy was inaugurated with the first direct detection of GWs from a binary black hole (BBH) merger by the Advanced Laser Interferometer Gravitational-wave Observatory (Advanced LIGO)~\cite{ref:Aasi2015} in 2015~\cite{ref:Abbott2016}. This groundbreaking discovery was followed by the first joint observation of GWs and electromagnetic counterparts from a binary neutron star (BNS) merger, achieved by Advanced LIGO, Advanced Virgo~\cite{ref:Acernese2015}, and other telescopes, paving the way for multimessenger astronomy~\cite{ref:Abbott2017_b}. Over the course of three observing runs (O1, O2, and O3), 90 GW events from compact binary coalescences (CBCs) were reported~\cite{ref:Abbott2019, ref:Abbott2020_a, ref:Abbott2021_a, ref:Abbott2021_b}. These events included two neutron star-black hole (NSBH) mergers~\cite{ref:Abbott2021_c} and two BNS mergers~\cite{ref:Abbott2017_a, ref:Abbott2020_b}. The detection of GWs, alongside electromagnetic waves and neutrinos from these mergers, is vital for understanding the physical properties of neutron star interiors, which are reflected in their equation of state. Now, with the commencement of the fourth observing run (O4) in May 2023, which includes the participation of KAGRA~\cite{ref:Akutsu2019}, expectations are high for more GW detections from binary systems with neutron stars.

Traditionally, GWs from CBC sources have been analyzed using the matched-filtering technique~\cite{ref:Allen2012} with theoretical approximants, phenomenological models, and templates derived from numerical simulations~\cite{ref:Creighton_2011,ref:Maggiore_2008}. In this technique, the signal-to-noise ratio (SNR) is computed by correlating the detector's strain data with each template in a large bank that covers a wide parameter space, taking into account variations in source masses and/or spins. This method, however, can be computationally intensive, especially for complex GW signals that incorporate elements such as higher-order modes, precession, or orbital eccentricity. This complexity underscores the need for more efficient algorithms to manage the growing volume of GW data.

In response to this challenge, deep learning approaches, particularly convolutional neural networks (CNNs), have been increasingly applied in the GW field. These applications range from parameter estimation of CBC sources~\cite{ref:Gabbard2021, ref:Dax2023} to sky localization~\cite{ref:Chatterjee2019, ref:Sasaoka2022, ref:Kolmus2022} and classification of transient noises~\cite{ref:Zevin2017, ref:Bahaadini2018, ref:Sakai2022}. The effectiveness of CNNs in detecting GWs from BBH mergers was first demonstrated in 2018 by \citeauthor{ref:George2018}~\cite{ref:George2018} and \citeauthor{ref:Gabbard2018}~\cite{ref:Gabbard2018}. These initial studies have since been expanded to include more sophisticated models that use real detector noise and account for various signal complexities like the spin effect, precession, higher-order modes, or eccentricity~\cite{ref:George2018_2, ref:Fan2019, ref:Gebhard2019, ref:Wang2020, ref:Wei2021_bbh, ref:Xia2021, ref:Morales2021, ref:Lopac2022, ref:Schafer2022_1, ref:Schafer2022_2, ref:Ma2022, ref:Schafer2023, ref:Ruan2023, ref:Nousi2023, ref:Murali2023}. There are also some studies targeting BNS~\cite{ref:Krastev2020, ref:Schafer2020, ref:Lin2021, ref:Krastev2021, ref:Wei2021_bns, ref:Baltus2021, ref:Baltus2022, ref:Aveiro2022} or NSBH signals~\cite{ref:Menendez2021, ref:Yu2021, ref:Wei2021_eccentric, ref:Qiu2023, ref:Garg2023}, which are more challenging than BBH signals due to their longer duration and smaller amplitude.

Two main types of CNNs have been employed in this research: one-dimensional (1D) CNNs, which process whitened time-series data, and two-dimensional (2D) CNNs, which analyze time-frequency maps. Although most studies have favored 1D CNNs~\cite{ref:George2018, ref:Gabbard2018, ref:George2018_2, ref:Gebhard2019, ref:Fan2019, ref:Wei2021_bbh, ref:Xia2021, ref:Schafer2022_1, ref:Schafer2022_2, ref:Ma2022, ref:Baltus2022, ref:Nousi2023, ref:Krastev2020, ref:Schafer2020, ref:Lin2021, ref:Krastev2021, ref:Yu2021, ref:Baltus2021, ref:Qiu2023, ref:Garg2023}, a subset has opted for 2D CNNs~\cite{ref:Morales2021, ref:Lopac2022, ref:Wei2021_bns, ref:Menendez2021, ref:Wei2021_eccentric, ref:Aveiro2022, ref:Murali2023}. 1D CNNs are preferred for their efficiency in not generating time-frequency maps, thereby reducing processing time. On the other hand, 2D CNNs excel at capturing the temporal evolution of GW frequencies in their input.
For the analysis of GWs from core-collapse supernovae, \citeauthor{ref:Iess2023}~\cite{ref:Iess2023} conducted a comparative study of 1D and 2D CNNs, alongside long short-term memory networks. Their approach involved combining these models by averaging their outputs. However, to our knowledge, a similar comprehensive comparison of various CNN architectures for CBC sources has not been extensively explored.

A common method for analyzing and interpreting CNNs involves pinpointing the segments of input data that significantly influence the model's predictions. This analysis can be performed using class activation mapping (CAM) techniques~\cite{ref:Zhou2015} or by assessing the contribution of each input feature to the model's output. In our prior research~\cite{ref:Sasaoka2023}, we applied CAM techniques to a CNN classifier designed for GWs from core-collapse supernovae. This investigation revealed that the model primarily focused on specific GW modes within the input spectrogram to make predictions.

In the current study, we train both 1D and 2D CNN models to detect and classify GWs from CBC sources. We then develop an ensemble model that combines these two CNN types. Our analysis includes a detailed comparison of the detection performance of these models across each  type of CBC signals. To distinguish the different aspects that 1D and 2D models focus on within the input, we employ the integrated gradients technique~\cite{ref:Sundararajan2017}. This approach allows us to identify the influential regions in the input that guide the models' predictions, revealing distinct areas of focus between the 1D and 2D models.

The paper is organized as follows. Section~\ref{sec:method} details our datasets, the architecture of the CNN models, and the theoretical background of the CNN analysis methods. Section~\ref{sec:result} presents the classification performance and a comprehensive analysis of our trained models. Finally, we conclude our findings in Sec.~\ref{sec:concl}.

\section{Method\label{sec:method}}
Our CNN models are trained to classify strains at three detectors LIGO Hanford (H1), LIGO Livingston (L1), and Virgo (V1) into four distinct classes: BBH, NSBH, BNS, and pure noise. This section provides a detailed description of the datasets used for both training and testing our models. Following this, we describe the architecture and training procedures of the CNN models. Lastly, we address the dimensionality reduction technique implemented in our study, as well as the methodology employed for computing feature attribution, which are crucial for interpreting the models' decision-making processes.
\subsection{Dataset}
To train and test our model, we used nonprecessing CBC signals and injected them into noise obtained from O3 real data at H1, L1, and V1, which are available at the Gravitational Wave Open Science Center~\cite{ref:Abbott2023}.
\subsubsection{Signal and noise generation}
To construct our datasets, non-precessing CBC signals were generated using the LIGO Algorithm Library Suite (LALSuite)~\cite{ref:Lalsuite}. Specifically, BBH signals were simulated using the \texttt{SEOBNRv4} approximant~\cite{ref:Bohe2017}, based on the effective-one-body method, while NSBH and BNS signals were generated using the \texttt{SpinTaylorT4} approximant~\cite{ref:Sturani2010}, a time-domain post-Newtonian model incorporating spin effects. For BBH signals, component masses were uniformly sampled in the range of 5 to 80 $M_\odot$. NSBH signals had NS masses sampled between 1 and 2 $M_\odot$, and BH masses between 5 and 35 $M_\odot$. The component masses of BNS signals ranged uniformly from 1 to 2 $M_\odot$. The individual components have spins aligned with the orbital angular momentum, uniformly distributed between 0 and 0.99. These waveforms were sampled at a rate of 4096 Hz. We used four-second data segments, with the merger event uniformly placed between 3.8 and 3.9 sec. Although NSBH and BNS signals are typically longer than 4 sec, we found this segment length is sufficient to discriminate between different classes. The use of shorter-segment signals also reduces the memory requirements for training models. The sky position of the source, defined by declination and right ascension, was randomly selected, and GW amplitude calculations were performed considering the antenna pattern functions and time delays across detectors. These computations utilized the PyCBC library~\cite{ref:Pycbc}.

For noise samples and background noise for signal samples, real strain data from GPS time 1238163456 to 1238659072 was used for the
training set, 1238663168 to 1239162880 was used for the validation set, and 1239166976 to 1239875584 was used for the test set. Data around the GW event time reported in the GWTC-2.1 catalog~\cite{ref:Abbott2021_a} were excluded.

\subsubsection{Preprocessing}
After the signal samples were truncated to four-second segments, they were scaled based on the computed optimal matched-filter SNR, defined as
\begin{equation}
    \rho = \sqrt{4\int_{f_\text{min}}^{f_\text{max}}\frac{|\tilde{h}(f)|^2}{S_n(f)}\mathrm{d}f},
\end{equation}
where $\tilde{h}(f)$ is the Fourier transform of the truncated signal and $S_n(f)$ is the one-side power spectral density of the noise, estimated using Welch's method~\cite{ref:Welch1967}. The integration was performed from a cutoff frequency of 20 Hz up to the Nyquist frequency. The training and validation signals were scaled so that the network SNR of the three detectors, given by
\begin{equation}
    \rho_{\mathrm{net}} = \sqrt{\rho_{\mathrm{H1}}^2+\rho_{\mathrm{L1}}^2+\rho_{\mathrm{V1}}^2},
\end{equation}
followed a uniform distribution between 8 and 24, while the SNRs of the test signals ranged from 3 to 24. After each signal was injected in noise, we whitened the sample in frequency domain using the power spectral density. For the input to the 2D CNN model, we generated time-frequency maps using the Q transform~\cite{ref:Chatterji2004} of the whitened samples, defined by
\begin{equation}
    X(\tau, \phi, Q) = \int_{-\infty}^{\infty}\tilde{x}(f+\phi)\tilde{w}^*(f, \phi, Q) e^{2\pi if\tau} \mathrm{d}f,
\end{equation}
where $Q$ is the quality factor, and the Connes window functions is used as the window function~\cite{ref:Chatterji2005}.

\begin{figure*}[tb]
    \centering
    \includegraphics[width=0.98\textwidth]{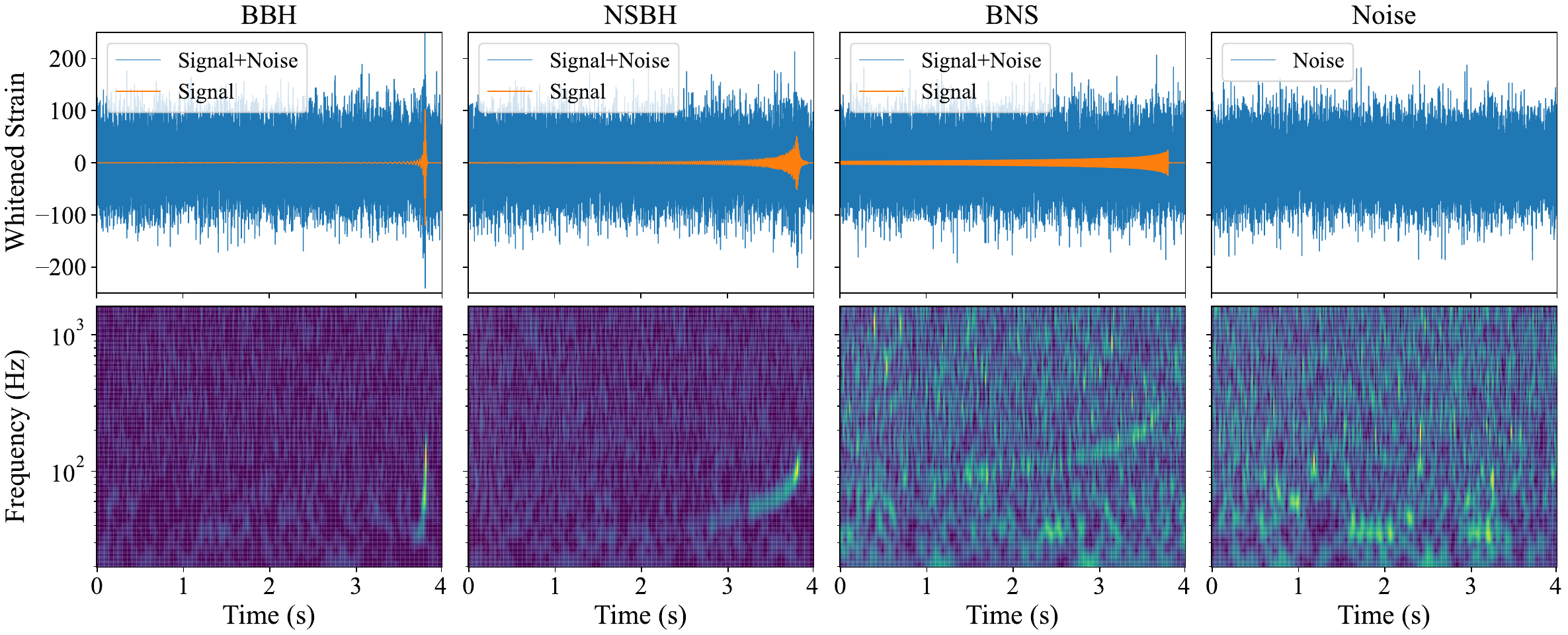}
    \caption{Example strain data at H1 detector in the training set. The upper figures show the whitened time-series data used as input to the 1D model, and the lower figures show the time-frequency maps for the 2D model. The component masses of the BBH signal are $51.3\ M_\odot$ and $50.9\ M_\odot$, whereas in the NSBH sample, the respective masses are $33.3\ M_\odot$ and $1.83\ M_\odot$, and the masses of the BNS sample are $1.54\ M_\odot$ and $1.40\ M_\odot$. The single-detector SNR of each signal is 15, and the merger time is fixed at 3.9 s.}
    \label{fig:sample}
\end{figure*}

The final dataset comprised 408,000 training samples, 408,000 validation samples, and 528,000 test samples. Each dataset had an equal distribution of 25\% BBH, 25\% NSBH, 25\% BNS, and 25\% pure noise samples. Representative samples from each class in the training set are displayed in Fig.~\ref{fig:sample}.

\subsection{Model}
\subsubsection{1D CNN}
One-dimensional CNN consists of 1D convolutional filters. Let $x^{c}_{i}$ be the $i$th value of the $c$th channel of the input series and $y^{m}_{i}$ be the $i$th value of the $m$th channel of the output. The output of a 1D convolutional filter is given by
\begin{equation}
y^{m}_{i} = \sum_{c=0}^{C-1} \sum_{k=0}^{K-1} w^{m}_{k}x^{c}_{i+k}+b^{m},
\end{equation}
where $K$ is the kernel size and $C$ is the number of input channels. Weight parameters $w$ and bias parameters $b$ are learned during training processes.

Our 1D CNN model takes a three-channel whitened time series at H1, L1, and V1 as input. Our implementation uses a 54-layer deep residual network (ResNet-54) which was proposed in Ref.~\cite{ref:Nousi2023}. ResNet is a type of deep CNN architecture that uses residual blocks to address the vanishing gradient problem commonly encountered in deep networks. It achieves this by adding \textit{skip connections} between layers, enabling the network to learn residual functions and make training deep networks more efficient~\cite{ref:He2016}. Details of the ResNet-54 model architecture can be found in Ref.~\cite{ref:Nousi2023}.

For input normalization, a deep adaptive input normalization layer~\cite{ref:Passalis2020} is employed in this analysis as used in Ref.~\cite{ref:Nousi2023} to address nonstationary noise that appears in real detector noise. In this layer, unlike conventional normalization, shifting and scaling parameters for normalizing the input are optimized during the training. Including the parameters in the normalization layer, the model has a total of 1,935,698 trainable parameters.

The employed architecture is one of the state-of-the-art models in BBH detection that surpassed the matched-filtering pipeline in a specific condition~\cite{ref:Nousi2023}.
Alternatively, we also employed the CNN architecture used in Ref.~\cite{ref:Qiu2023} designed to detect all types of CBC signals, however, it did not show a better performance than the ResNet-54 model in our datasets.

\subsubsection{2D CNN}
Two-dimensional CNN consists of 2D convolutional filters. Let $x^{c}_{i,j}$ be the $(i,j)$ component of the $c$th channel of the input image and $y^{m}_{i,j}$ be the $(i,j)$ component of the $c$th channel of the output. The output of a 2D convolutional filter is given by
\begin{equation}
y^{m}_{i,j} = \sum_{c=0}^{C-1} \sum_{k=0}^{K-1} \sum_{l=0}^{L-1} w^{m}_{k,l}x^{c}_{i+k, j+l}+b^{m}
\end{equation}
where $(K, L)$ is the kernel sizes and $C$ is the number of input channels. Weight parameters $w$ and bias parameters $b$ are learned during training processes.

The ResNet-50 model~\cite{ref:He2016}, a variant of ResNet with proven efficacy in image recognition tasks, forms the basis of our 2D CNN. This model processes three-channel images of time-frequency maps and includes 23,508,548 trainable parameters. We adopt this model because it is one of the most widely used 2D CNNs in GW signal detection and has a similar number of layers to our 1D model. Its efficiency has been validated in previous studies~\cite{ref:Menendez2021, ref:Wei2021_bns, ref:Wei2021_eccentric}.
\subsubsection{Ensemble model}
In our ensemble approach, we combine the outputs of the 1D and 2D CNN models to enhance predictive performance. This is achieved by first training a fully connected neural network, which takes as input a concatenated vector of features extracted from the trained 1D and 2D models. The input vector, with a dimension of 10240, is processed through a hidden layer of 200 units, outputting a four-dimensional vector. The network incorporates a Leaky ReLU layer~\cite{ref:Maas2013} and a dropout layer~\cite{ref:Hinton2012} with a 0.25 dropout rate for regularization. The ensemble network comprises 2,049,004 trainable parameters.

\begin{figure}[tb]
    \centering
    \includegraphics[width=0.99\columnwidth]{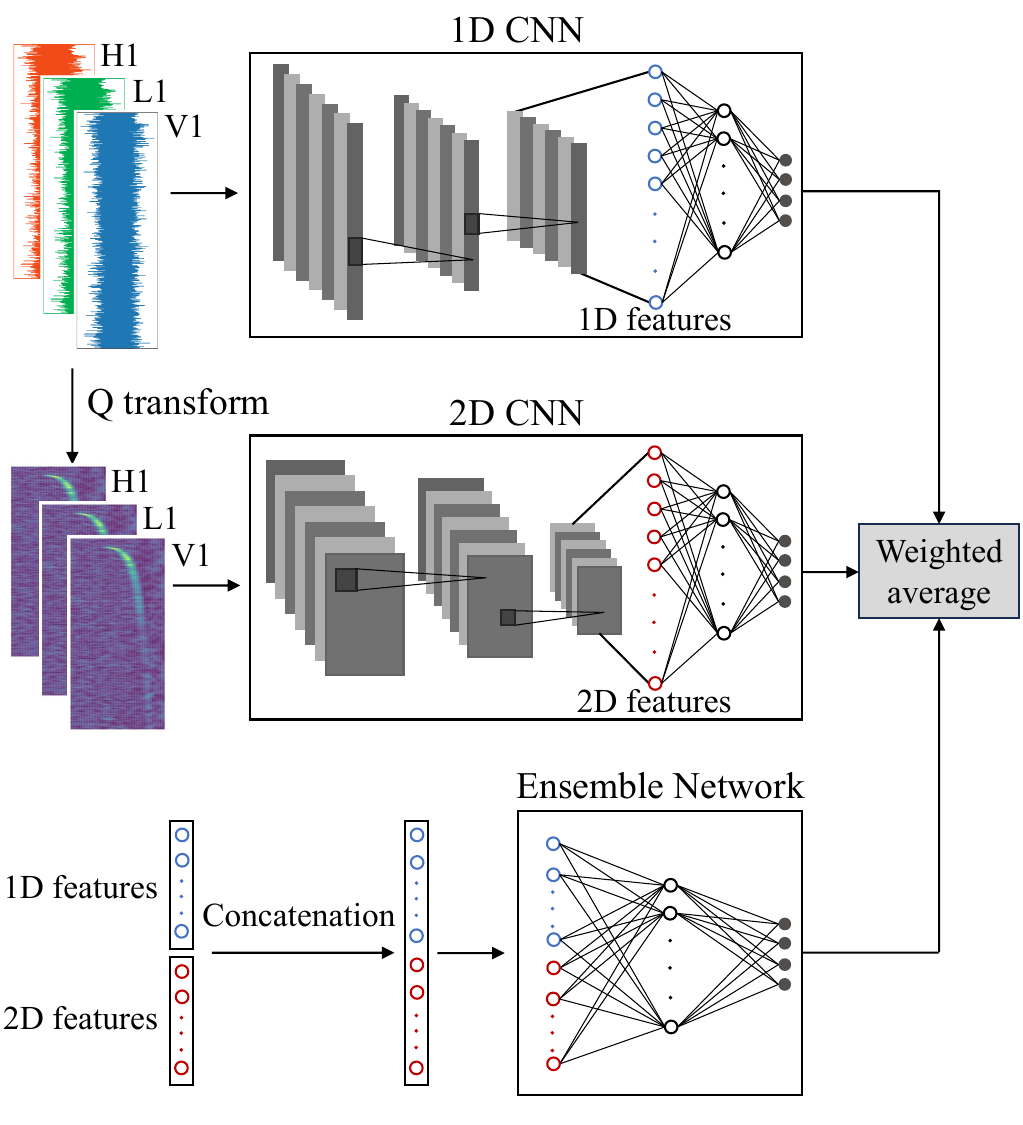}
    \caption{Illustration of the ensemble model. The final output is the weighted average of the outputs from the 1D CNN, 2D CNN, and ensemble network.}
    \label{fig:ensemble_flow}
\end{figure}

For the final model output, we employ a weighted average of the predictions from the 1D, 2D, and ensemble network. The weights, optimized for accuracy on the validation set, are set at 0.4 for each of the 1D and 2D CNNs and 0.2 for the ensemble network. The ensemble model is illustrated in Fig.~\ref{fig:ensemble_flow}.
\subsubsection{Training process}
Both the 1D and 2D CNN models, as well as the ensemble network, were developed using the PyTorch library~\cite{ref:Pytorch} and trained on four NVIDIA Tesla V100 GPUs. All models were trained using categorical cross entropy as the loss function and Adam optimizer~\cite{ref:Kingma2014} with an initial learning rate of $10^{-3}$. The learning rate was controlled by PyTorch's \texttt{ReduceLROnPlateau} method. During the training of the 1D and 2D CNNs, we implemented the curriculum learning technique~\cite{ref:Bengio2009}. This method involves initially training with high SNR samples and progressively incorporating lower SNR samples, thereby improving learning efficiency and model performance.

The 1D CNN underwent 300 epochs of training with a minibatch size of 1024. In contrast, the 2D CNN was trained for 45 epochs using a minibatch size of 36. The ensemble network's training lasted for 25 epochs with a minibatch size of 256.

\subsection{t-distributed stochastic neighbor embedding}
To understand the ability of 1D and 2D convolutional filters to extract meaningful features from the input for classification, we analyzed feature maps from the final convolutional layer. Given the high-dimensional nature of these feature maps, we utilized the t-distributed stochastic neighbor embedding (t-SNE) technique~\cite{ref:van2008} for dimensionality reduction. 

In the original space, the conditional probability $p_{j|i}$ that a data point $x_i$ would pick a data point $x_j$ as its neighbor is modeled as a Gaussian distribution centered at $x_i$, defined as
\begin{equation}
    \label{eq:conditional_p} p_{j|i} = \frac{\exp(-\|x_i-x_j\|^2/2\sigma_i^2)}{\sum_{k\neq i}\exp(-\|x_i-x_k\|^2/2\sigma_i^2)},
\end{equation}
where $\sigma_i$ is the standard deviation. We define the joint probability $p_{ij}$ as the symmetrized conditional probabilities, which can be expressed as $p_{ij} = (p_{j|i} + p_{i|j})/2n$, where $n$ is the number of data points. In the low-dimensional space, the Student t distribution with one degree of freedom, defined by
\begin{equation}
    q_{ij} = \frac{(1+\|y_i-y_j\|^2)^{-1}}{\sum_{k\neq l}(1+\|y_k-y_l\|^2)^{-1}},
\end{equation}
is used to quantify the similarity of data points. The optimal low-dimensional representations are obtained by minimizing the Kullback-Leibler divergence of the distributions $p_{ij}$ and $q_{ij}$, given by
\begin{equation}
    C = \sum_{i,j} p_{ij}\log \frac{p_{ij}}{q_{ij}}.
\end{equation}
The value of $\sigma_i$ in Eq.~(\ref{eq:conditional_p}) is determined by selecting the hyperparameter called perplexity, which is defined as 2 to the power of the Shannon entropy. The perplexity can be interpreted as a measure of the number of valid neighbors, and typical values are between 5 and 50~\cite{ref:van2008}. We set the perplexity at 25.
\subsection{Integrated gradients}
To discern which aspects of the inputs significantly influence the predictions in our trained 1D and 2D models, we employed the integrated gradients method~\cite{ref:Sundararajan2017}. While class activation mapping techniques~\cite{ref:Zhou2015, ref:Selvaraju2017} are commonly used for such analysis, they often yield low-resolution saliency maps, especially in deep models. To circumvent this limitation, the integrated gradients method provides high-resolution feature attribution maps, proving advantageous for our analysis.

The integrated gradients method is grounded in two axioms that attribution methods should satisfy: (i) \textit{sensitivity}, where any difference in one feature between the input and the baseline resulting in different predictions should receive a non-zero attribution, and (ii) \textit{implementation invariance}, where the attributions for two functionally equivalent networks should be always identical.

Consider a function $F$ that represents a network and let $x$ be the input and $x'$ be the baseline input. The feature attribution map is calculated by examining the path from the baseline $x'$ to the input $x$ and accumulating the network's gradients along this path. A point on this path can be expressed as $x'+\alpha (x-x')$ where $\alpha$ varies from 0 to 1. The integrated gradients along the $i$th dimension for an input $x$ are defined as
\begin{equation}
    \mathrm{IG}_i(x) = (x_i-x_i')\int_0^1 \frac{\partial F(x'+\alpha(x-x'))}{\partial x_i}\mathrm{d}\alpha.
\end{equation}
In practice, this integration is approximated using the Riemann sum, described as
\begin{equation}
    \mathrm{IG}_i(x) \approx (x_i-x_i')\sum_{k=1}^N \frac{\partial F(x'+\frac{k}{N}(x-x'))}{\partial x_i}\frac{1}{N}.
\end{equation}
Here $N$ represents the number of interpolation steps. For accurately approximating the integral, a step size ranging from 20 to 300 is typically effective~\cite{ref:Sundararajan2017}. In our implementation, we chose $N=30$ steps. The Captum library~\cite{ref:Kokhlikyan2020} was utilized to compute the attribution maps using the integrated gradients method.
\section{Results and Discussion\label{sec:result}}
\subsection{Model performance}
\begin{figure}[tb]
    \centering
    \includegraphics[width=0.98\columnwidth]{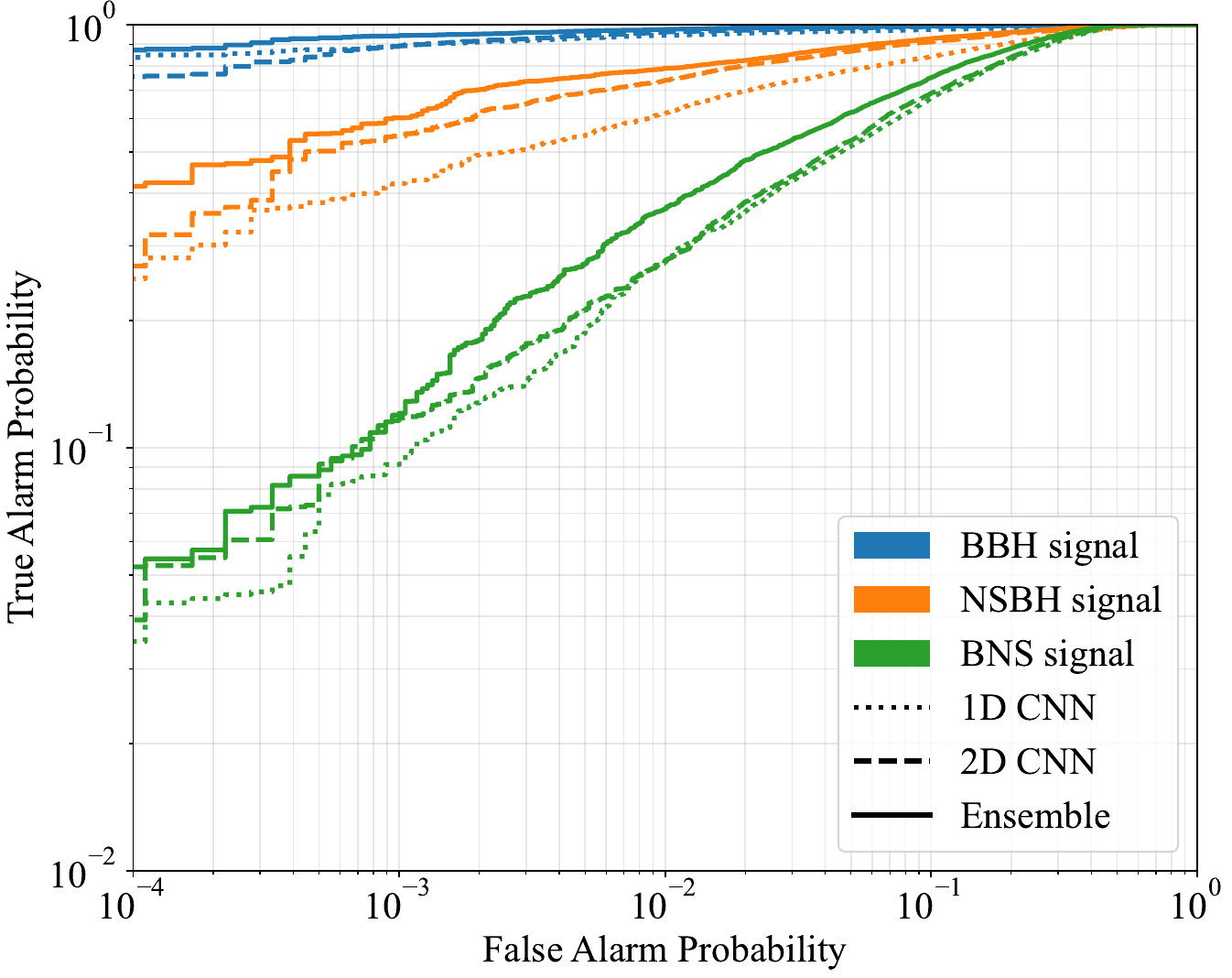}
    \caption{ROC curves of the three models for BBH, NSBH, and BNS signals at a fixed network SNR of 8. The SNRs are computed with four-second signals.}
    \label{fig:roc}
\end{figure}

To evaluate the performance of our three models (1D, 2D, and the ensemble model), we first examined the receiver operating characteristic (ROC) curves for each signal type. The ROC curve plots the true alarm probability (TAP) against the false alarm probability (FAP) at various classification thresholds. As depicted in Fig.~\ref{fig:roc}, the ROC curves for each type of GW signal at a fixed network SNR of 8 show distinctive sensitivities.

It was observed that all models exhibited the highest sensitivity to BBH signals, followed by NSBH and BNS signals. This trend aligns with expectations considering the relative amplitude of each signal type. Notably, the ensemble model demonstrated superior performance across all signal types. For BBH signals, the performance ranking was ensemble, followed by the 1D and then the 2D model. In contrast, for NSBH and BNS signals, the 2D model outperformed the 1D model. This variation in performance can be attributed to the transient nature of BBH signals, which are more effectively captured by the 1D convolution in time-series data. Conversely, the smaller amplitudes of NSBH and BNS signals, which are more challenging to identify in time-series data, render the 2D model more effective. This difference highlights the effectiveness of combining the 1D and 2D models.

\begin{figure}[tb]
    \centering
    \includegraphics[width=0.98\columnwidth]{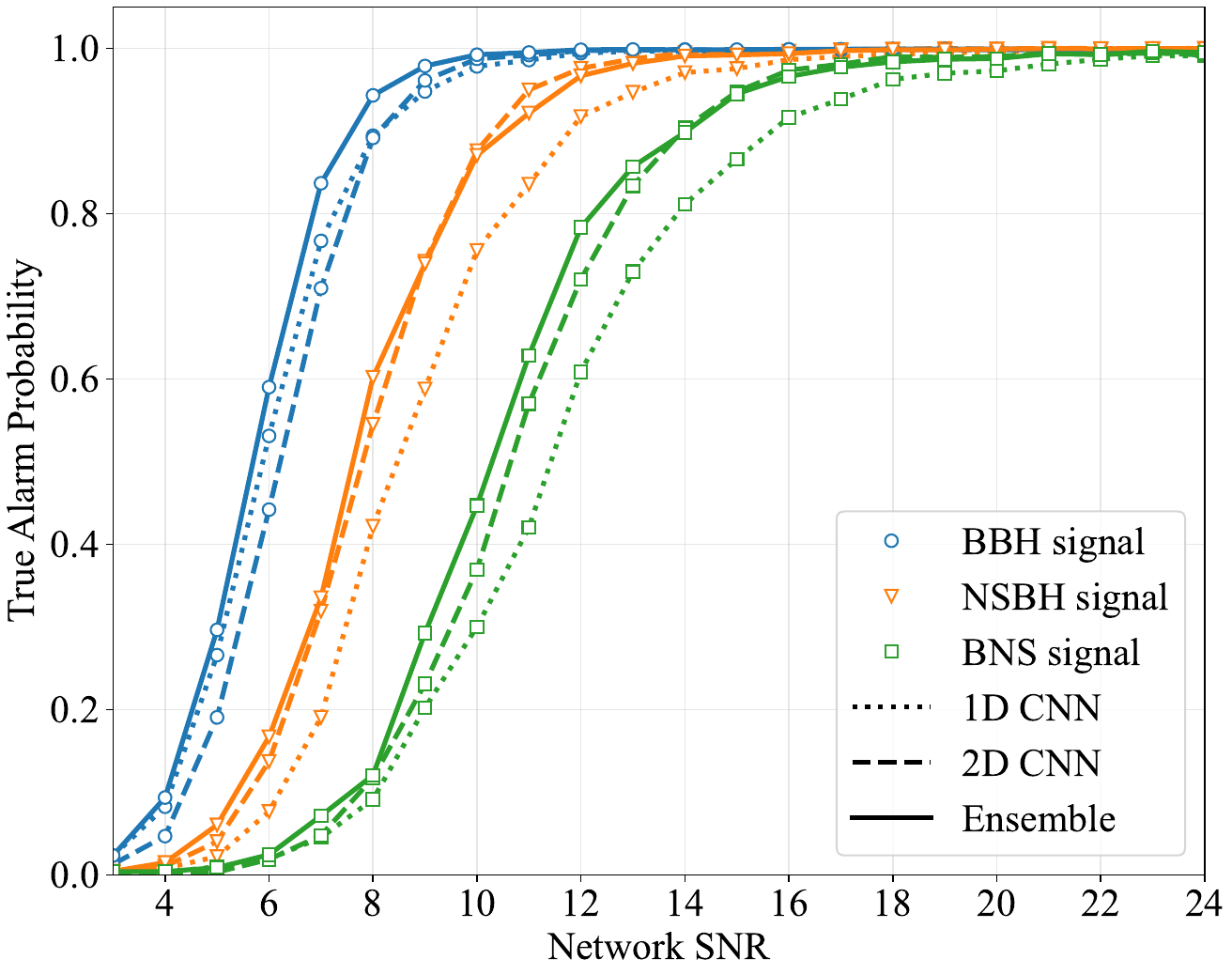}
    \caption{Sensitivity curves of the three models for BBH, NSBH, and BNS signals at a fixed false alarm probability of 0.001. The SNRs are computed with four-second signals.}
    \label{fig:tpr}
\end{figure}

We further calculated the detection sensitivity for each signal type as a function of network SNR. Figure~\ref{fig:tpr} shows the sensitivity curves for the three models at a fixed FAP of 0.001. The 1D model's sensitivity is on par with that reported in Ref.~\cite{ref:Qiu2023}, where the model was trained using single-detector input. Their model's sensitivity saturates at a single-detector SNR of $\rho_{\text{L1}}\geq 8$ for BBH signals, at $\rho_{\text{L1}}\geq 10$ for NSBH signals, and at $\rho_{\text{L1}}\geq 13$ for BNS signals. Our 1D model, however, reaches saturation for BBH signals at $\rho_{\text{net}}\geq 12$, for NSBH signals at $\rho_{\text{net}}\geq 17$, and for BNS signals at $\rho_{\text{net}}\geq 22$. Given that the network SNR of three detectors is roughly $\sqrt{3}$ times that of a single-detector SNR, the performance of our 1D model is consistent with their model. The ensemble model further enhances this performance, lowering the saturation SNRs to 10 for BBH signals, 14 for NSBH signals, and 21 for BNS signals.
\subsection{Feature map}
\begin{figure}[tb]
    \centering
    \includegraphics[width=0.98\columnwidth]{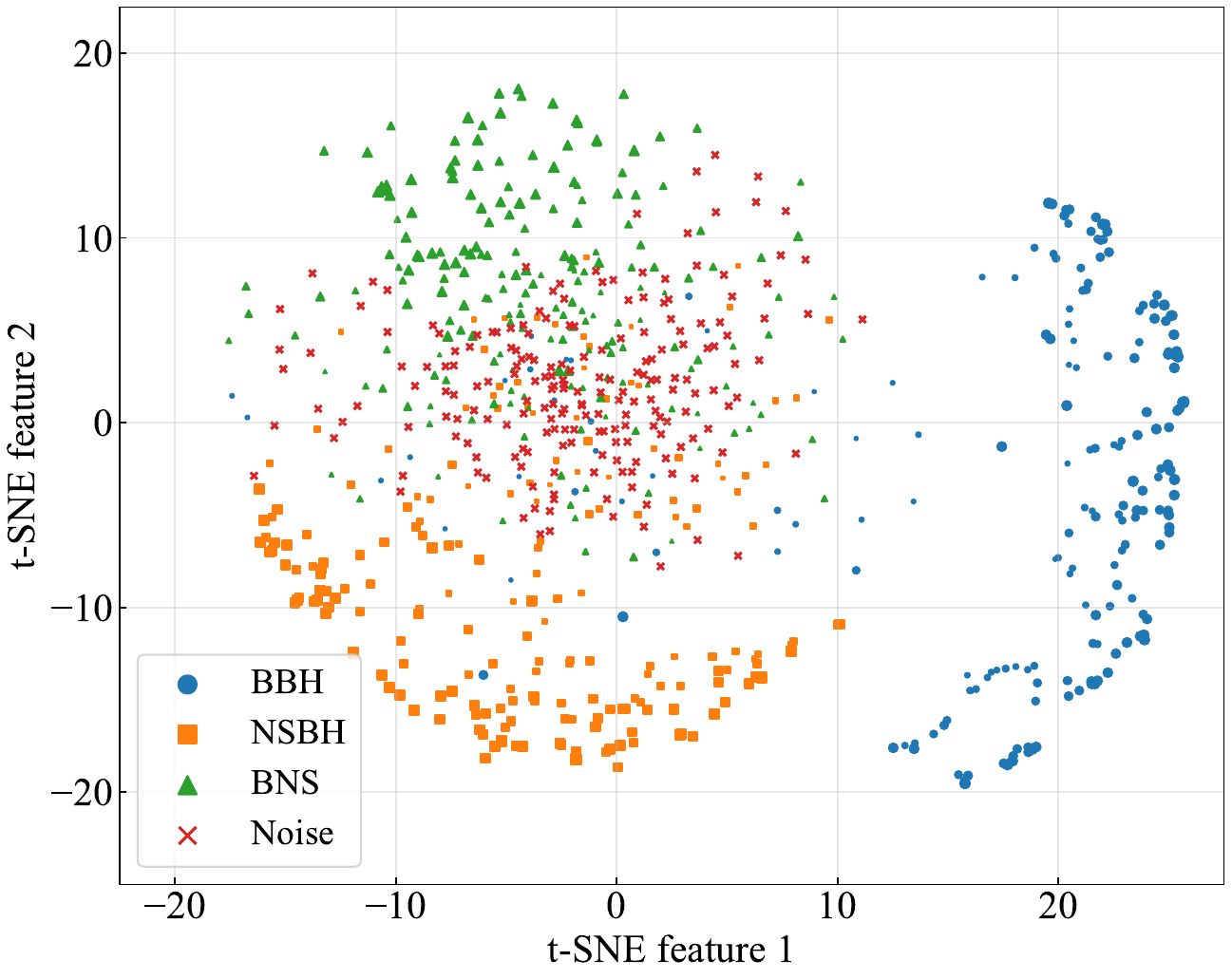}
    \caption{Two-dimensional representations of feature maps of the 1D model by t-SNE. The marker size of signal sample corresponds to the SNR.}
    \label{fig:t_sne_1d}
\end{figure}

\begin{figure}[tb]
    \centering
    \includegraphics[width=0.98\columnwidth]{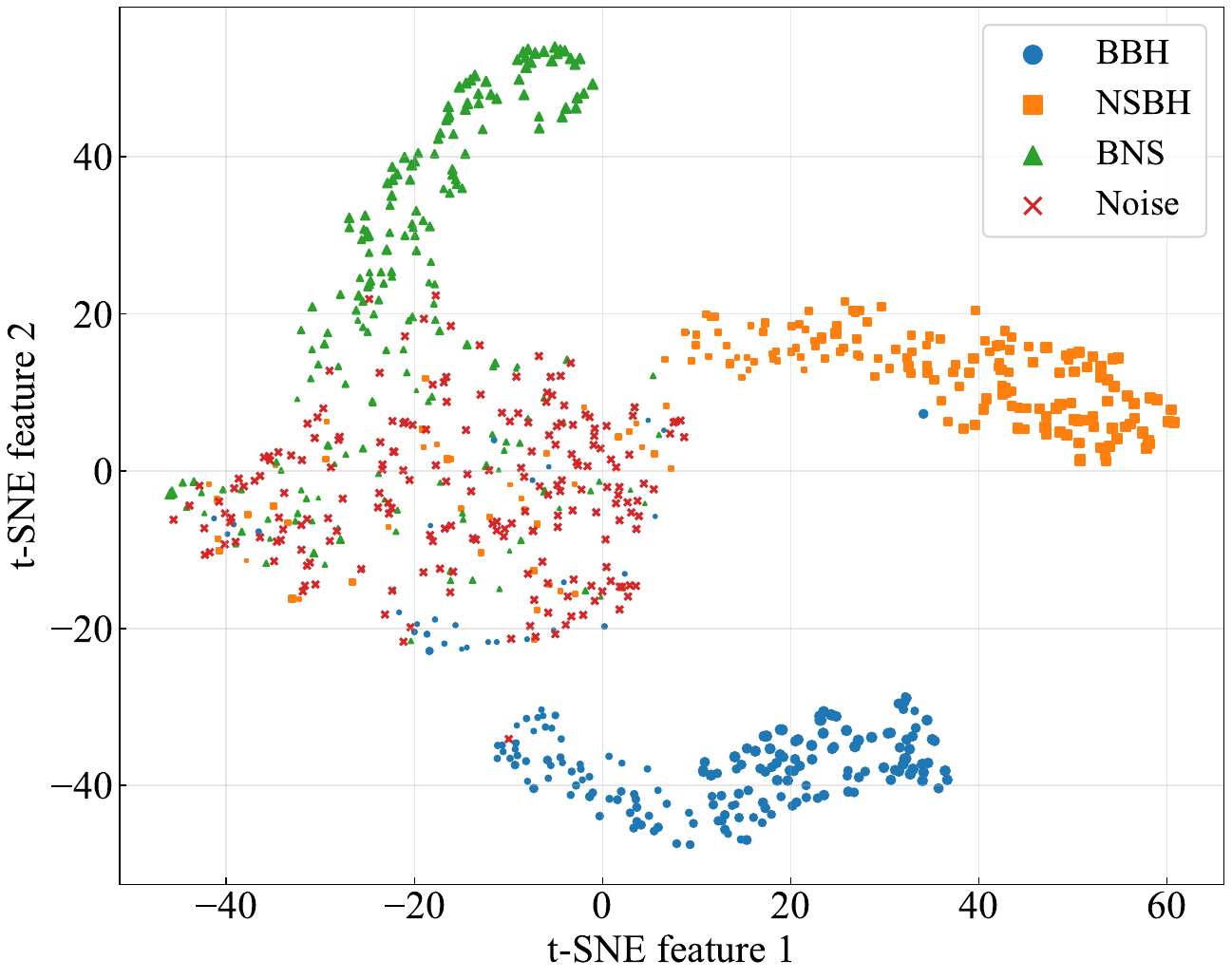}
    \caption{Two-dimensional representations of feature maps of the 2D model by t-SNE. The marker size of signal sample corresponds to the SNR.}
    \label{fig:t_sne_2d}
\end{figure}

We extract feature maps from the final convolutional layers. Since these vectors are fed into fully connected layers to make predictions, these feature maps represent the characteristics of each class. Figures~\ref{fig:t_sne_1d} and \ref{fig:t_sne_2d} display the t-SNE projections of the feature maps for the trained 1D and 2D models, respectively. For these visualizations, we randomly selected 200 samples from the test set. In the figures, the size of each marker representing a signal sample is proportional to its SNR. Smaller markers indicate lower SNR signals, while larger markers correspond to higher SNR signals.

Figure~\ref{fig:t_sne_1d} shows that the high-SNR BBH and NSBH samples are distinctly separated from the noise cluster, indicating effective classification of these signals by the 1D model. However, BNS samples are observed to be closer to the noise cluster, suggesting less clear differentiation for this signal type. Low-SNR signals across all types are more diffusely distributed within the noise cluster. In contrast, Fig.~\ref{fig:t_sne_2d} indicates that the 2D model has an improved ability to separate not only the BBH and NSBH signals but also the BNS signals from the noise cluster. This indicates that the 2D model may be more proficient at identifying features of BNS signals compared to the 1D model.

Additionally, we explored embedding the feature maps into a three-dimensional space. However, this analysis revealed similar characteristics to those observed in the two-dimensional embeddings.
\subsection{Attribution map}
To understand how input features contribute to model predictions, we generated attribution maps for each type of signal using the integrated gradients method. For this analysis, a single signal was randomly selected from the test set, and ten distinct noise samples were added to create ten different input samples. Attribution maps were produced for each input sample, and their average was computed to discern universal characteristics of the attribution.

\begin{figure}[tb]
    \centering
    \includegraphics[width=0.98\columnwidth]{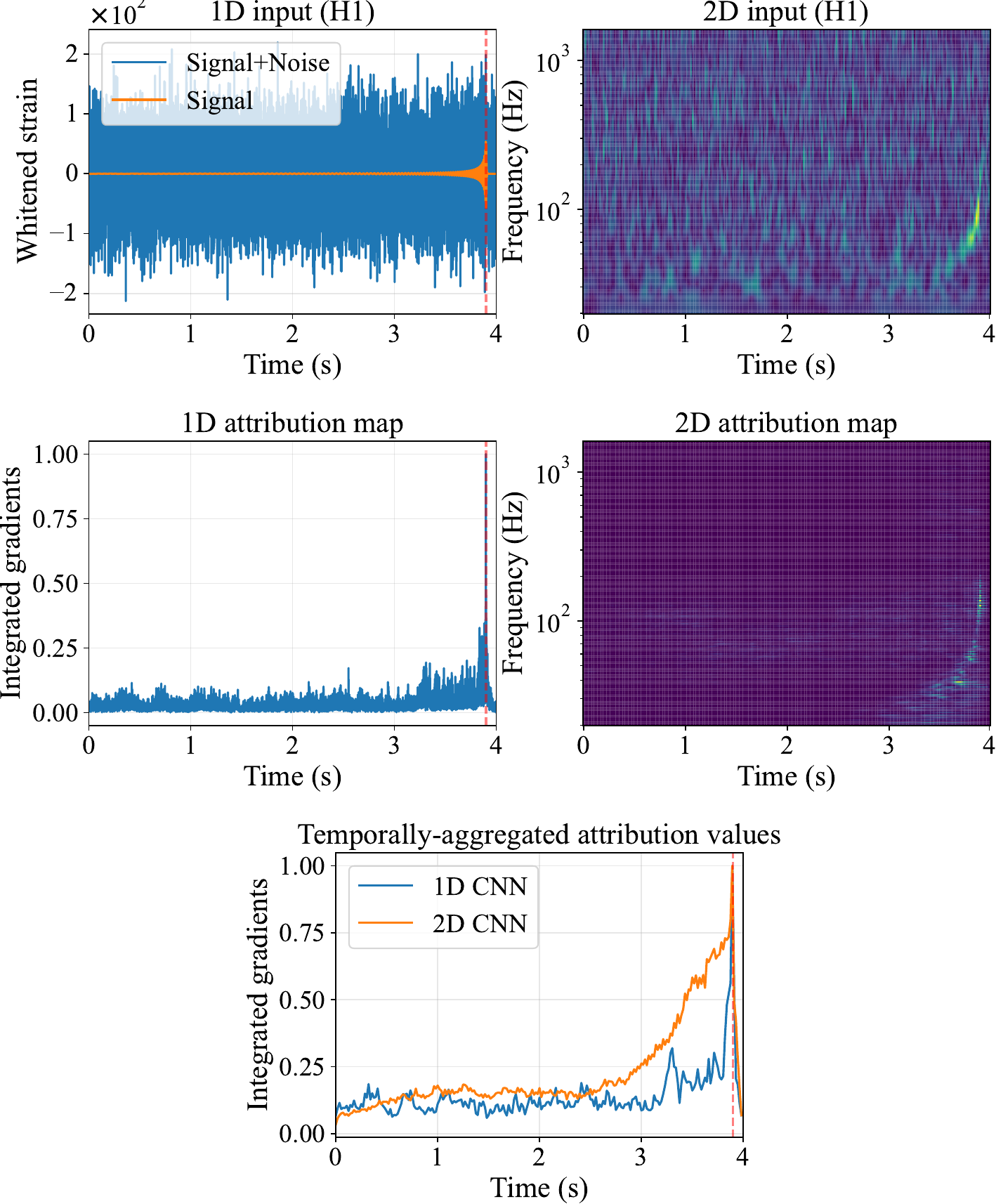}
    \caption{Attribution maps of the 1D and 2D models for a BBH signal and corresponding input samples at H1 detector. The component masses are $37.7\ M_\odot$ and $6.94\ M_\odot$. The red dashed line shows the time of coalescence. The bottom plot shows the values of integrated gradients summed over all frequencies for each time bin.}
    \label{fig:ig_bbh}
\end{figure}

\begin{figure}[tb]
    \centering
    \includegraphics[width=0.98\columnwidth]{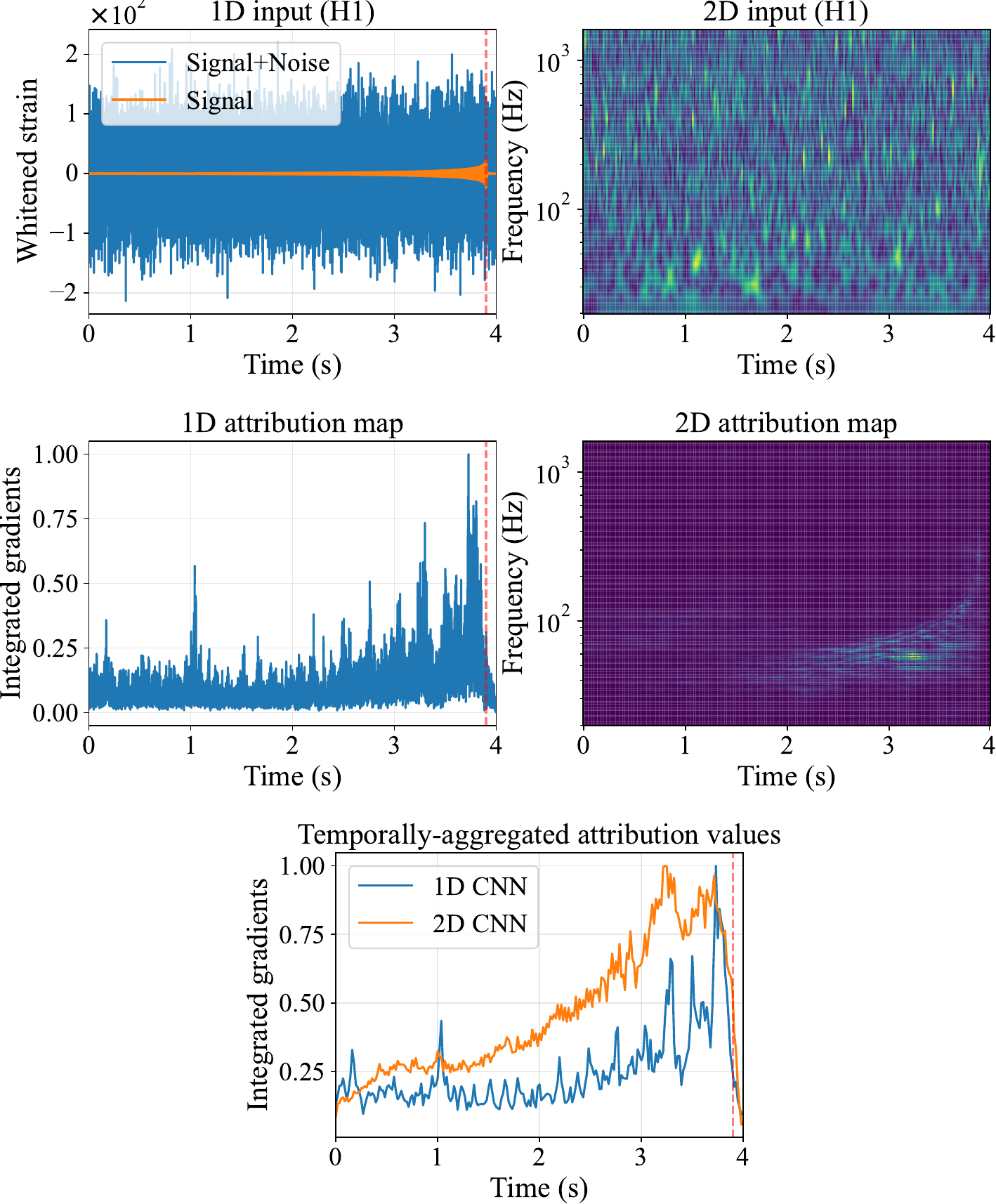}
    \caption{Attribution maps of the 1D and 2D models for a NSBH signal and corresponding input samples at H1 detector. The component masses are $6.78\ M_\odot$ and $1.99\ M_\odot$. The red dashed line shows the time of coalescence. The bottom plot shows the values of integrated gradients summed over all frequencies for each time bin.}
    \label{fig:ig_nsbh}
\end{figure}

\begin{figure}[tb]
    \centering
    \includegraphics[width=0.98\columnwidth]{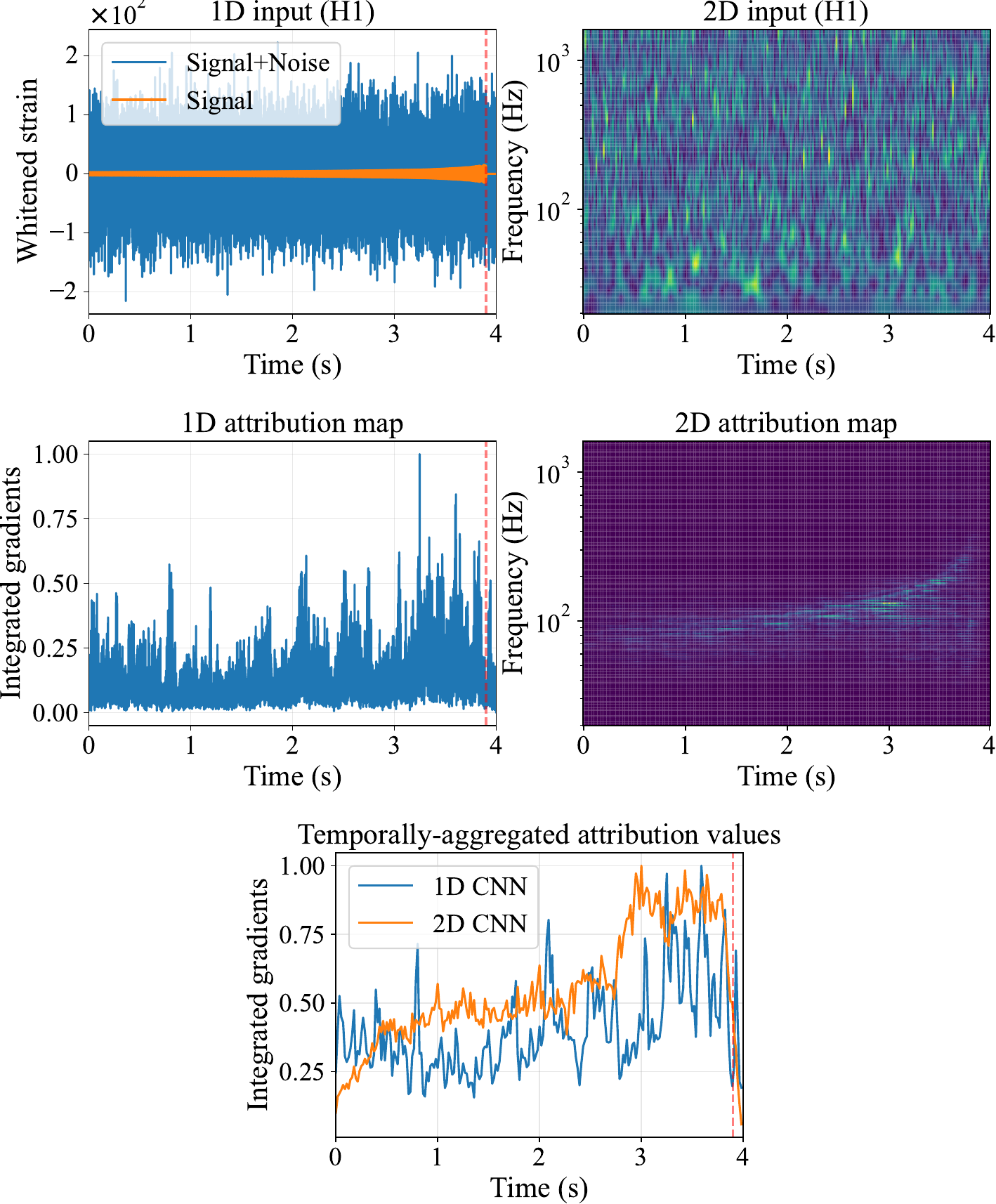}
    \caption{Attribution maps of the 1D and 2D models for a BNS signal and corresponding input samples at H1 detector. The component masses are $1.18\ M_\odot$ and $1.15\ M_\odot$. The red dashed line shows the time of coalescence. The bottom plot shows the values of integrated gradients summed over all frequencies for each time bin.}
    \label{fig:ig_bns}
\end{figure}

Figures~\ref{fig:ig_bbh}, \ref{fig:ig_nsbh}, and \ref{fig:ig_bns} present the attribution maps for a BBH, NSBH, and BNS signal, respectively, as identified by the 1D and 2D models. Each signal had a fixed network SNR of 20, and the attribution maps for both models were normalized to a [0, 1] range. The bottom plots in Figs.~\ref{fig:ig_bbh}, \ref{fig:ig_nsbh}, and \ref{fig:ig_bns} show the values of integrated gradients summed over all frequencies for each time bin for 1D and 2D model, respectively. The values are normalized to a maximum integrated gradients of one for each model.

In Fig.~\ref{fig:ig_bbh}, the 1D model shows significant contributions from data at the coalescence time of the BBH signal. In contrast, the 2D model’s attribution map indicates that the 2D model focuses on the entire inspiral signal. Both the 1D and 2D models exhibit similar characteristics when integrated-gradients values are temporally aggregated, however, the 2D model sees data at more broader time frame than the 1D model.

As for the NSBH signal, the 1D model exhibits multiple peaks in the integrated gradients values before the coalescence, with the peak values progressively increasing, shown in Fig.~\ref{fig:ig_nsbh}. Since the model not only detects the signal but also classifies it into three classes, the data prior to the time of coalescence seem to be more significant than the data at the time of coalescence for determining that the signal is NSBH, not BBH. In the 2D model, the feature contribution of the entire inspiral is large, as in the case of the BBH signal. Similar characteristics of the NSBH sample are seen in the BNS sample in Fig.~\ref{fig:ig_bns}, but in the case of 1D attribution map of the BNS sample, peaks are also seen at earlier times and the overall values of the integrated gradients are generally identical. This indicates that the model focuses on various parts of the input time series data, which is reasonable because the inspiral signal is longer than the other signals. The 2D attribution map of the BNS sample shows that the 2D model accurately captures the BNS chirp signal on the input spectrogram, and it demonstrates the consistent performance of the 2D model for BNS events. The temporally aggregated attribution maps have similar characteristics for each signal type for the 1D and 2D models, but the 2D model shows a greater emphasis on longer signal durations than the 1D model.

In summary, from the attribution maps, we observe that the 1D model places greater emphasis on the time preceding coalescence, especially as the waveform lengthens, with significant contributions from specific moments in the inspiral phase. Conversely, the 2D model assesses the entire chirp waveform in the spectrogram, classifying based on the shape of the chirp, i.e., the temporal evolution of its frequency.
\section{Conclusions\label{sec:concl}}
In this study, we explored the application of three distinct models: a 1D CNN, a 2D CNN, and an ensemble model for detecting and classifying GWs from CBC sources. The 1D model, trained on whitened time-series data, excelled in identifying BBH signals, while the 2D model, trained on Q-transformed spectrograms, showed superior performance with NSBH and BNS signals. Overall, the ensemble model demonstrated the most robust classification capability across all signal types.

The effectiveness of combining 1D and 2D models was further reinforced through feature map visualization using the t-SNE technique and attribution map analysis via the integrated gradients method. We observed that the 1D model tends to focus on data preceding the merger time, especially as signal duration increases. In contrast, the 2D model scrutinizes the entire chirp waveform, capturing the intricacies of GW signals more comprehensively. These differences in focus and performance between the models highlight the benefits of their integration. 

While our study presents results based on a specific selection of architectures for 1D and 2D CNNs, it is important to recognize that the field offers a wide variety of CNN architectures. Future research exploring additional architectures may provide a more comprehensive understanding and validation of the conclusions drawn in this study.

We discussed the performance of the models by fixing a FAP at 0.001, but given that each data sample is 4 s long, this would result in roughly one false positive every hour, making our method insufficient for real application. Efficiency could be enhanced by incorporating a subsequent model, such as a binary classifier differentiating BBH signals from noise, to further reduce the FAP. Validation is also required to address unbalanced data, considering the source population.

Our classification models hold potential for analyzing long continuous data through a sliding window approach. Although our models were trained to identify GW signals occurring between 3.8 and 3.9 sec within four-second segments, sliding input window with a step size, for example, of 0.1 seconds, allows us to detect signals at any time point in principle. However, this approach may lead to encountering multiple triggers within a single event, requiring further tuning, which will be addressed in future studies.
\begin{acknowledgments}
The authors are grateful to Koustav Chandra for useful comments. This research was supported in part by the Japan Society for the Promotion of Science (JSPS) Grant-in-Aid for Scientific Research [No. 22H01228 (K.\ Somiya), and Nos. 19H01901, 23H01176 and 23H04520 (H.\ Takahashi)].
This research was also supported by the Joint Research Program of the Institute for Cosmic Ray Research, University of Tokyo and Tokyo City University Prioritized Studies. 
This research has made use of data or software obtained from the Gravitational Wave Open Science Center~\cite{ref:gwosc}, a service of the LIGO Scientific Collaboration, the Virgo Collaboration, and KAGRA. This material is based upon work supported by NSF's LIGO Laboratory which is a major facility fully funded by the National Science Foundation, as well as the Science and Technology Facilities Council (STFC) of the United Kingdom, the Max-Planck-Society (MPS), and the State of Niedersachsen/Germany for support of the construction of Advanced LIGO and construction and operation of the GEO600 detector. Additional support for Advanced LIGO was provided by the Australian Research Council. Virgo is funded, through the European Gravitational Observatory (EGO), by the French Centre National de Recherche Scientifique (CNRS), the Italian Istituto Nazionale di Fisica Nucleare (INFN) and the Dutch Nikhef, with contributions by institutions from Belgium, Germany, Greece, Hungary, Ireland, Japan, Monaco, Poland, Portugal, Spain. KAGRA is supported by Ministry of Education, Culture, Sports, Science and Technology (MEXT), Japan Society for the Promotion of Science (JSPS) in Japan; National Research Foundation (NRF) and Ministry of Science and ICT (MSIT) in Korea; Academia Sinica (AS) and National Science and Technology Council (NSTC) in Taiwan. 

\end{acknowledgments}

\bibliography{apssamp}

\end{document}